\documentstyle[preprint,aps]{revtex}

\def\hef{ $^4$He }
\def\het{ $^3$He }
\def\rr { {\bf r } }
\def\rp { {\bf r' } }
\def\rpar { {\bf \displaystyle r_{\parallel}} }
\def\bk { {\bf k} }
\def\dis{ \displaystyle}

\begin{document}
\draft

\title{Structural and dynamical properties of superfluid helium:\\
	a density functional approach.}

\author{    F. Dalfovo$^a$, A. Lastri$^a$, L.Pricaupenko$^b$,
            S. Stringari$^a$, and J. Treiner$^b$ }

\address{$^a$ Dipartimento di Fisica, Universit\`a di Trento, \\
              38050 Povo, Italy }
\address{$^b$ Division de Physique Th\'eorique, Institute de
Physique Nucl\'eaire, \\ 91406  Orsay Cedex, France  }

\date{\today}
\maketitle

\begin{abstract}

     We present a novel density functional for liquid  $^4$He, properly
accounting for the static response function and the phonon-roton
dispersion in the uniform liquid. The functional is used to study both
structural and  dynamical properties of  superfluid helium in various
geometries. The equilibrium properties of the free surface,  droplets
and films at zero temperature are calculated. Our predictions agree
closely to the results of {\it ab initio} Monte Carlo calculations, when
available. The introduction of a phenomenological velocity dependent
interaction, which accounts for backflow effects, is discussed. The
spectrum of the elementary excitations of the free  surface and films
is studied.

\end{abstract}

\pacs{PACS number:67.40}

\narrowtext

\section{Introduction}
\label{sec:intro}

    The aim of the present work is to present a  density functional  theory,
which treats static and dynamic properties of liquid helium on the same
ground, and is accurate at the  microscopic level.

    The investigation of the properties of liquid helium in different
geometries  has a long story. Extensive work has been devoted to the search for
and the understanding of superfluid effects in finite  systems like helium
droplets, or in quasi two dimensional systems like  helium films on solid
substrates, or within porous materials. Liquid helium  exhibits very peculiar
properties, such as the propagation  of surface modes in the short wavelength
regime,  quantum evaporation  of atoms produced by rotons impinging on the
surface,  wetting  and prewetting transitions on solid substrates. The
nucleation of  bubbles at negative pressure and of  quantized vortices near
walls are further examples of interesting  phenomena where both the
inhomogeneity of the liquid and quantum correlations play an important
role. These phenomena, among others, make  liquid helium particularly appealing
from  the viewpoint of quantum many  body theories. Theory and experiments,
however, have not yet  a  satisfactory overlap. One difficulty comes from the
fact that {\it ab initio} calculations  are  still  hard to implement
for inhomogeneous systems.  On the other hand, phenomenological theories, which
are quite  successful in describing macroscopic properties, are  not always
adequate enough to investigate the behaviour of the system on the scale  of
interatomic distances.  A major progress in this direction has been recently
made in the framework of density functional theory (DFT).

    Several density functionals have been developed in
the  last years for applications to quantum fluids. The method consists of
writing the energy of the  many body system as a functional of the one-body
density;  from a given  functional one extracts the equilibrium state,
by minimizing  the energy with  respect to  the density, as well as the
excited states, by  solving time-dependent equations of motion.  An
accurate phenomenological  density functional for liquid \hef
(Orsay-Paris functional) has been recently introduced \cite{Dup90}; it has
proven to be quite reliable in different contexts, such as helium films
and wetting phenomena \cite{Pav91,Treiner,Cheng1,Cheng2,Cheng3},
vortices in bulk liquid \cite{Dal92} or droplets \cite{Dal94}. For
dynamical properties it gives predictions which are close to the
Feynman approximation for elementary excitations. In particular it
does not account for backflow effects \cite{Pri94}.

    In the present work we present a new functional. We follow the same
ideas which lead to the Orsay-Paris functional, taking a similar  two body
interaction of Lennard-Jones type and including phenomenologically short
range  correlations. The main difference with respect to the Orsay-Paris
functional is the addition of two new terms which allow for a  better
description of both static and dynamic properties on the scale of the
interatomic spacing. These terms are:

\begin{itemize}
\item  a non local term depending on gradients of the density, which
allows to reproduce the experimental static response function in
a wide range of wave vectors, as well as its pressure dependence;

\item  a term depending on local variations of the velocity field
(backflow effects), which allows to reproduce the experimental
phonon-roton dispersion in bulk.
\end{itemize}

	The velocity dependent term has a form similar to the one already
introduced by Thouless \cite{Tho69}, who studied the flow of a dense
superfluid. The idea is to model backflow effects, which are
important at small wavelengths, by a non-local kinetic energy term.

With these ingredients the predictions for several properties of non
uniform systems are significantly improved with respect to the ones
given by  previous functionals.  The accuracy of the density functional
theory is now comparable to the one of {\it ab initio} Monte Carlo
calculations. The resulting approach, although phenomenological,
represents a powerful and  accurate tool whenever {\it ab  initio}
calculations become hard to implement. It  allows one to investigate a
wide variety of systems of different sizes and in different geometries,
from few atoms to the bulk liquid, with limited numerical
efforts.

	The paper is organized as follows:  in Sect.~\ref{sec:statics}
we introduce the density functional for static calculations. We
emphasize and motivate the differences with respect to the Orsay-Paris
functional. The results for the equilibrium properties of the free
surface, droplets and films are given in Sect.~\ref{sec:equiconf}.
In Sect.~\ref{sec:dynamics} we discuss the
application of the density functional theory to dynamics, showing the
connection with the Feynman approximation and with the formalism of the
Random Phase Approximation. We introduce a phenomenological
current-current interaction which makes the density functional
quantitative in the description of the spectrum of  excited states. We
discuss the role of this new term using sum rule arguments. Finally, in
Sect.~\ref{sec:excited} we present the results  for the excited states of
the free surface and of films.

\section{Ground state calculations}
\label{sec:statics}

 	In the density functional formalism at zero temperature, the energy
of a Bose system is assumed to be a functional of a complex macroscopic wave
function $\Psi$:
\begin{equation}
     E \  = \   \int \!d\rr \ {\cal H} [\Psi, \Psi^*] \ \ \ .
\label{eq:e}
\end{equation}
The function $\Psi$ can be written in the form
\begin{equation}
     \Psi (\rr,t) = \Phi(\rr,t) \exp \left({i \over \hbar} S (\rr,t)
\right)  \ \ \ .
\label{eq:psi}
\end{equation}
The real function $\Phi$ is related to the diagonal one-body density by
$\rho=\Phi^2$, while the phase $S$ fixes the velocity of the
fluid  through the relation  ${\bf v} = (1/m) \nabla S$.
In the calculation of the ground
state, only states with zero velocity  are considered, so that the energy
is simply  a functional of the one-body density $\rho(\rr)$. A natural
representation is given by
\begin{equation}
E \ = \ \int \! d\rr \ {\cal H}_0 [\rho] \ =  \  E_c[\rho] \ + \
\int \! d\rr  \ {\hbar^2 \over 2m}  ( \nabla \sqrt{\rho})^2  \ \ \ ,
\label{eq:ec}
\end{equation}
where the second term on the r.h.s. is a quantum pressure, corresponding
to  the kinetic  energy of a Bose gas of non uniform density. The quantity
$E_c[\rho]$ is  a "correlation energy"; it incorporates the effects of
dynamic correlations induced by the interaction.

  Ground state configurations are obtained by minimizing the energy of
the system with respect to the density. This leads to the Hartree-type
equation
\begin{equation}
\left\{ - {\hbar^2 \over 2m} \nabla^2  + U [\rho, \rr] \right\}
\sqrt{\rho(\rr)} \ = \ \mu \sqrt{\rho(\rr)} \ \ \ ,
\label{eq:hartree}
\end{equation}
where $U[\rho, \rr] \equiv \delta E_c /\delta \rho(\rr)$ acts as a  mean
field,  while the chemical potential $\mu$ is introduced in order to
ensure the  proper normalization  of the density to a fixed number of
particles.

	For a weakly interacting Bose system  the expression of ${\cal H}_0$
can be derived on a rigorous basis, yielding the  well known
Gross-Pitaevskii theory \cite{Gro61}. Since liquid helium is a strongly
correlated system, such a derivation, starting from first principles, is not
available. One  then resorts to approximate schemes for the correlation
energy. Krotscheck and coworkers  \cite{Kro85,Kro92,Cle93}, for
instance,   have developed  a variational hypernetted chain/Euler-Lagrange
(HNC/EL)  theory in  which the one-body effective potential $U$ is
evaluated  using the microscopic Hamiltonian.
An alternative approach consists of writing  a
phenomenological expression for the  correlation energy, whose  parameters
are fixed to reproduce  known  properties of the bulk liquid.

	A simple functional was introduced in  Ref.~\cite{Str87a,Str87b} to
investigate properties of the free surface and droplets of both \hef  and
\het. The correlation energy, in  analogy with the formalism of zero-range
Skyrme interactions in nuclei \cite{Bri72}, was written as
\begin{equation}
E_c[\rho] = \int d\rr  \left[ {b\over 2} \rho^2 + {c\over 2}
\rho^{2+\gamma} + d (\nabla \rho)^2 \right] \ \ \ ,
\label{eq:skyrme}
\end{equation}
where $b,c$ and $\gamma$ are phenomenological parameters fixed   to
reproduce the ground state energy, density and compressibility of the
homogeneous liquid at zero pressure, and $d$ is adjusted to the surface
tension of the liquid. The first two terms correspond  to a local
density  approximation  for the correlation energy, and non local
effects  are included through the gradient correction. The local density
approximation is currently  used to describe exchange-correlation
energy in electron systems; its use  for liquid helium  does not
provide quantitative results \cite{Ji86}.
Non-local terms, like $(\nabla \rho)^2$ in  (\ref{eq:skyrme}),
turn  out to be crucial for the description of inhomogeneous
liquid helium. Non-locality effects have been included in DFT in a more
realistic  way by Dupont-Roc et al. \cite{Dup90}, who generalized
Eq.~(\ref{eq:skyrme}) accounting for the finite range of the atom-atom
interaction. The resulting functional
has proven to be reliable in describing several inhomogeneous systems
\cite{Dup90,Pav91,Treiner,Cheng1,Cheng2,Cheng3,Dal92,Dal94}.

In the present work we follow the spirit of Ref.~\cite{Dup90}. We  use
a similar two-body finite range interaction screened at short distances,
and a weighted density (or "coarse grained" density) to account for
short range correlations.
The most important feature  of this approach is that the static
response function of the liquid can be reproduced even at finite wave
vectors $q$, up to the roton region. The static  response  function
$\chi(q)$ fixes the linear  response of the system to  static density
perturbations and is a key quantity in density functional theories.
It is easily calculated from functional  (\ref{eq:ec}) by taking
the second functional derivative of the energy  in $q$-space
\cite{note:chi}:
\begin{equation}
- \chi^{-1}(q) = {\hbar^2 q^2 \over 4 m} + {\rho \over V}
\int \! d\rr d\rp { \delta E_c \over \delta \rho(\rr) \delta
\rho (\rp) } e^{- i {\bf q} (\rr -\rp) } \ \ \ .
\label{eq:secder}
\end{equation}
A major advantage of liquid helium  is that $\chi(q)$ is known
experimentally. It is related to the
inverse  energy weighted moment of the dynamic  structure  function
$S(q,\omega)$, measured in neutron scattering, through the relation
\begin{equation}
\chi(q) \ = \ -2\, m_{-1} (q) \ \ \  ,
\label{eq:chi}
\end{equation}
the $n$-moment of $S(q,\omega)$ being defined as
\begin{equation}
m_n(q) \ = \int_0^{\infty} \! d\omega \ S(q,\omega) (\hbar \omega)^n
\ \ \ .
\label{eq:moments}
\end{equation}
The experimental data for $\chi(q)$ in the uniform liquid at zero
pressure \cite{Cow71} are shown in Fig.~\ref{fig:chi} (circles). The
prediction of the zero-range functional of Ref.~\cite{Str87a} (i.e., with
the correlation energy given in Eq.~(\ref{eq:skyrme})) and of the finite
range Orsay-Paris functional \cite{Dup90} are also shown as dotted and
dashed lines respectively. The  $q=0$ limit is fixed by the
compressibility of the system,  which  in both cases is an input of the
theory. It ensures the correct  behaviour in the long wavelength limit
and, consequently, the  correct description of systems characterized
by smooth density variations as happens, for example, in the free surface
or in helium droplets. The height of the peak  of the static response
function in the roton region,  $q \simeq 2$ \AA$^{-1}$,  is
important in characterizing  structural properties on the interatomic
length  scale (for instance the layered structure of helium films).
Therefore, the first important idea is to improve on the
Orsay-Paris functional in order to better reproduce the experimental peak
of $\chi(q)$ in the roton region.

In the present work, the function ${\cal H}_0$, entering functional
(\ref{eq:ec}), is taken in the form
\begin{eqnarray}
{\cal H}_0 &=& \ {\hbar^2 \over 2m} (\nabla \sqrt{\rho})^2 \ + \
        {1\over 2} \int \!  d\rp \ \rho(\rr) V_l(|\rr-\rp|) \rho(\rp)
	\ + \ {c_2 \over 2} \rho(\rr)  (\bar \rho_{\rr})^2
        \ + \ {c_3  \over 3} \rho(\rr)  (\bar \rho_{\rr})^3
\nonumber
\\    	&-&    {\hbar^2 \over 4m} \alpha_s \int \! d\rp \ F(| \rr -\rp |)
	       \left(1-{\tilde \rho(\rr) \over \rho_{0s}} \right)
	       \nabla \rho(\rr) \cdot \nabla \rho(\rp)
	       \left(1-{\tilde \rho(\rp) \over \rho_{0s}} \right)
\ \ \ .
\label{eq:h0}
\end{eqnarray}
	The first term on the r.h.s. is the quantum pressure, as in
Eq.~(\ref{eq:ec}).  The second term contains a two-body interaction $V_{l}$,
which is the  Lennard-Jones interatomic potential, with the standard
value of the hard core radius of 2.556 \AA \ and of the well-depth 10.22
K, screened at
short distance  ($V_l \equiv 0$ for $r<h$, with $h=2.1903$\AA).  The  weighted
density  $\bar \rho$ is the average of $\rho (\rr)$ over a sphere  of
radius $h$: \begin{equation}
\bar \rho (\rr) = \int \! d\rr \  \Pi_h (|\rr-\rp|) \rho(\rp) \ \ ,
\label{eq:rhobar}
\end{equation}
	where $\Pi_h(r) =3 (4 \pi h^3)^{-1}$ for $r \le h$ and zero elsewhere.
The two terms containing $\bar \rho$, with the parameters   $c_2
=-2.411857 \times 10^4$ K \AA$^6$ and  $c_3=1.858496 \times 10^6$  K
\AA$^9$, account phenomenologically for short range correlations. All
these terms have a form similar to the Orsay-Paris functional
\cite{Dup90}. The Lennard-Jones potential is here screened in a  simpler
way, avoiding the fourth power for $r<h$, and the  dependence on $\bar
\rho$ is slightly different. The effects of these minor changes will be
discussed later. The last term in Eq.~(\ref{eq:h0}) is a completely new
term; it depends on the gradient of the density at different points and
corresponds to a non-local correction to the kinetic energy.  The function $F$
is a simple gaussian $F (r)= \pi^{-3/2} \ell^{-3} \exp(-r^2/\ell^2) $
with $\ell=1$ \AA, while $\alpha_s=54.31$ \AA$^{3}$. The parameters are fixed
to reproduce the peak of the static response  function in the bulk liquid. The
latter can be easily calculated by using  Eq.~(\ref{eq:secder}). One finds
\begin{eqnarray}
 - \chi^{-1} (q)  & = &  \frac{\hbar^2 q^2}{4m} + \rho \hat{V}_l (q)
               + c_2 (2 \hat{\Pi}_h (q) + \hat{\Pi}^2_h (q) )
               \rho^2 \nonumber\\
              & + & 2 c_3 (\hat{\Pi}_h (q) + \hat{\Pi}_h^2 (q) ) \rho^3  \ \
	      -{\hbar^2 \over 2 m} \alpha_s \rho (1-{\rho \over \rho_{0s}})^2
              q^2 \exp \left( - \frac{q^2 \ell^2}{4} \right) \ ,
\label{eq:chi-1}
\end{eqnarray}
	where $\hat{V}_l (q)$ and $\hat{\Pi}_h(q)$ are the Fourier transforms
of the  screened Lennard-Jones potential and the weighting function
$\Pi_h$,  respectively, while $\rho$ is the bulk density. The resulting
curve for the liquid at zero pressure ($\rho=\rho_0=0.021836$ \AA$^{-3}$)
is shown in Fig.~\ref{fig:chi} (solid line).  The factor $(1- \tilde \rho
/\rho_{0s})$, with $\rho_{0s}=0.04$ \AA$^{-3}$, is included in order to
obtain a pressure dependence of the static response function close to the
one predicted by Diffusion Monte Carlo simulations \cite{Mor92}. For
instance in the liquid near solidification ($\rho =0.02622$ \AA$^{-3}$) one
finds a peak of $\chi(q)$ about $10$ \% higher than at zero pressure, and
displaced by 0.1 \AA$^{-1}$ to larger  wavelengths.  Finally, the
quantity $\tilde \rho(\rr)$ is again a weighted  density, calculated using
the gradient-gradient interaction function $F$  as a weighting function:
\begin{equation}
\tilde \rho (\rr) = \int \! d\rp F(|\rr -\rp|) \rho(\rp) \ \ \ .
\label{eq:rhotilde}
\end{equation}
	Actually the density $\tilde \rho(\rr)$ is very close to the particle
density  $\rho(\rr)$ except in strongly inhomogeneous situations (like
helium adsorbed on a graphite substrate). For this reason one can safely
replace $\tilde \rho(\rr)$  with $\rho(\rr)$ for the investigation of the
free surface, helium droplets and films on weak binding substrates.
Stronger  constraints on the form of $F$ should be provided by the study of the
liquid-solid phase  transition.

	In a uniform liquid of constant density the energy per particle,  from
functional (\ref{eq:ec},\ref{eq:h0}), reduces to the power law
\begin{equation}
{ E \over N} = {b \over 2} \rho + {c_2\over 2} \rho^2 +  {c_3\over 3}
\rho^3 \ \ \  ,
\label{eq:eovern}
\end{equation}
where $b=-718.99$ K\AA$^3$ is the integral of the screened Lennard-Jones
potential $V_l$. Pressure and compressibility can be derived directly
by taking the first and second derivative of the energy:
\begin{equation}
P = \rho^2 {\partial \over \partial \rho} {E \over N}
\ \ \ \ ; \ \
- \chi^{-1}(0) = {\partial P \over \partial \rho}
\ \ \ .
\label{eq:pandchi}
\end{equation}
The experimental
values of the density, energy per particle and  compressibility of the
uniform system at zero pressure are used as input  to fix the parameters
$h, c_2$ and $c_3$. The resulting equation of state and the sound
velocity, $c^2=-[m\chi(0)]^{-1}$, are shown in  Figs.~\ref{fig:eqstate}
and \ref{fig:sound} respectively. The comparison  with
the results of Monte Carlo simulations  \cite{Bor94}  and with
experimental data  \cite{Abr70,DeB87}  shows that the present density
functional theory describes correctly the ground state of the bulk liquid
\hef at all pressures. Comments about the equation of state in the limit
of a quasi two-dimensional liquid  will be given when discussing the
structural properties of films.

 To conclude this section we emphasize again the main idea. The density
functional (\ref{eq:ec},\ref{eq:h0}) incorporates the correct long
range behaviour of the interatomic potential, and accounts for short
range correlations in a phenomenological way. The ground state of the
uniform liquid  is well reproduced at all pressures and, furthermore,
the response to small static density perturbations, up to the roton wave
length,  is also correctly reproduced. Significant differences with
respect to previous functionals are  expected in the predictions of
properties which depend on the behaviour of the fluid on the interatomic
length scale. Some  interesting examples are discussed in the following
section.

\section{Results for the equilibrium configuration of free surface,
droplets  and films}
\label{sec:equiconf}

\subsection{Free surface}
\label{sec:free}

	To compute the density  profile and the energy of a planar free
surface of \hef at zero temperature one has to solve the  integro-differential
Hartree equation (\ref{eq:hartree}) with the mean field $U$ extracted
from  (\ref{eq:h0}).  Both $\rho$ and $U$ depend only on the
coordinate orthogonal to the surface, so that the equation is
one-dimensional.  The integrals on parallel coordinates in the  non-local
terms of the functional can be written analytically.  The numerical
solution of non-linear Eq.~(\ref{eq:hartree}) is obtained by means of  an
iterative procedure and provides the density $\rho(z)$, from which the
surface tension $\sigma$ can be calculated through
\begin{equation}
\sigma = \int \! dz \ \left\{ {\cal H}_0[\rho] - \mu \rho
\right\}  \ \ \ .
\label{eq:surfenergy}
\end{equation}
The bulk density is kept fixed to the experimental value at zero pressure,
$\rho_0=0.021836$ \AA$^{-3}$. The  density profile is shown in
Fig.~\ref{fig:free}, together with the results given by the Skyrme
functional of Eq.~(\ref{eq:skyrme}) and the Orsay-Paris one. The
surface tension is  practically the same in the three cases.
The present functional gives  $\sigma=0.272$ K
\AA$^{-2}$, to be compared with the experimental values quoted in the
literature: $0.275$ K \AA$^{-2}$  \cite{Guo71} and $0.257$ K \AA$^{-2}$
\cite{Iin85}. The 10\%-90\% surface thickness is approximately $6$ \AA.
The value $7.6$ \AA,  extracted from X-ray scattering data \cite{Lur92},
is slightly larger, being closer to the result of the zero-range
functional of Eq.~(\ref{eq:skyrme}). However, the experimental value
is expected to depend on the form of the density profile used to fit
the measured reflectivity.

Notice that the density $\rho(z)$ resulting from expression
(\ref{eq:h0}) is not as smooth as the ones given by previous
calculations. It exhibits small oscillations which  appear
as shoulders on the  surface profile and  asymptotically die
in the bulk region. Oscillations of this type were predicted long
time  ago by Regge \cite{Reg72}.
In his theory the surface was treated as  a source of
elementary excitations producing static ripples on the  density profile.
The form of the ripples is connected to the  behaviour of the static
response function. The pronounced peak at the roton wavelength is
associated with the tendency of atoms to localize in "soft sphere
close packing". This tendency is opposed by the zero point motion of the
surface, whose thickness is larger than the interparticle
distance, and by compressibility effects, so that the density
oscillations turn out to be quite small. The role of
the static response function in characterizing the form of the surface
profile is clearly seen on Figs.~\ref{fig:chi}
and \ref{fig:free}. Of course one expects that functionals
describing the short wavelength  behaviour of $\chi(q)$ in a correct way
will give rise to more reliable predictions for the properties
of the fluid in the microscopic scale.  This will be explicitly
confirmed in the case of helium droplets, where our predictions can be
compared with Monte Carlo simulations.

\subsection{Droplets}
\label{sec:droplets}

The solution of the Hartree equation (\ref{eq:hartree}) in spherical
symmetry with a fixed number of particles $N$ provides the ground state
of \hef droplets. Since the density functional approach
is not time-consuming, it allows to compute the energy and the density profile
of droplets in a wide range of sizes. The  density
$\rho(r)$  for droplets with $8 \le N \le 60$ is shown  in  Fig.~\ref{fig:3d}.
Again one finds ripples on the surface profile; they are more pronounced than
in the case of the free surface, since the droplets have a size of the order
of a few interatomic distances and the "soft sphere close packing"
tends to produce shell structures.  This effect was already
suggested by Rasetti and Regge \cite{Ras78}, but subsequent theoretical
calculations  \cite{Str87b,Pan83,Sin89,Wha90} did not predict any clear
and systematic  oscillation in $\rho(r)$.  Only recently, sizable oscillations
have been found in the Diffusion Monte Carlo (DMC) calculations by Chin
and  Krotscheck \cite{Chi92}.   An example is given in Fig.~\ref{fig:clu70},
where we show the density profile of a droplet with $70$ particles. The solid
line is the result of the present density functional calculation, while the
DMC results of Ref.~\cite{Chi92} are represented by triangles.  The DMC data
exhibit more pronounced oscillations, but  the presence of   long-lived
metastable states, slowing  down the convergence in the  Monte Carlo algorithm,
cannot be completely ruled out \cite{Chi92}.  Recently the same  authors have
found oscillations in $\rho(r)$ even with a variational  approach based on the
HNC approximation  \cite{Chi94} (dashed line). Even though the HNC method
underestimates the central density, it predicts oscillations with amplitude
and phase in remarkable agreement with the ones of density functional theory.
An even
better agreement is found in the most recent DMC calculations by Barnett and
Whaley \cite{Wha94} (circles), where  the statistical error is significantly
reduced with respect to  Ref.~\cite{Chi92}.

	The detailed structure of $\rho(r)$, though interesting from a
theoretical viewpoint, is not yet observable experimentally with enough
accuracy to distinguish between a smooth profile and a profile with small
oscillations. It is thus important to explore the effects of the  "soft
sphere close packing" on the energy systematics, since the latter is
related to the mass distribution of droplets in the experimental beams
\cite{Toe90}.  In Fig.~\ref{fig:cluenergy} the energy per particle is
given as a function of $N$. The results of the present work (full  line)
are compared with the ones of previous functionals (dashed and dotted
lines), as well as with Monte Carlo simulations (dots from
Ref.~\cite{Chi92} and crosses from Ref.~\cite{Mel84}). First we note that
the accuracy of the density functional theory, compared with "ab initio"
simulations, increases progressively, following the  improvement in the
prediction of the static response function $\chi(q)$ in the
microscopic  region.
The agreement between the results of our functional  and Monte Carlo data
is  excellent. Second, as the energy is a smooth function of $N$, helium
clusters behave essentially as liquid droplets. Indeed the energy can be
easily fitted with a liquid drop formula:
\begin{equation}
{E \over N} = a_v
+ a_s N^{-1/3} + a_c N^{-2/3} + a_0 N^{-1} \ \ \  ,
\label{eq:liqdrop}
\end{equation}
where the volume coefficient $a_v$ is the chemical potential
in bulk,  the surface energy $a_s$ is fixed by the surface tension,
while $a_c$ and $a_0$ can be taken as fitting parameters.  The energy
calculated with the density functional
differs from the liquid drop fit by less than $0.02$ K  for all
droplets with $N> 30$.  This seems to rule out  apparently any shell
effect in the energy systematics.  However, the  relevant quantity
to investigate in this context is the
evaporation energy $[E(N-1)-E(N)]$. The latter is not as smooth as the
energy per particle. Figure \ref{fig:evap} shows the evaporation
energy predicted by the density functional (solid line) and  the
one obtained with the liquid drop formula (\ref{eq:liqdrop}) (dashed
line).  The difference between the two curves is also shown
(circles). When the difference is positive the droplets are more
stable than it is predicted by the liquid drop formula. We note
clear oscillations, having decreasing amplitude and increasing periodicity
as a function of $N$. The same kind of oscillations appear in the central
density of the droplets, as seen in  Fig.~\ref{fig:3d}. Since the distance
between two crests  of the surface ripples is practically constant  and
the droplet radius goes  approximately like $N^{1/3}$, the period of
oscillations of the central density as a function of $N$, as well as the
one of the evaporation energy,  increases as $N^{1/3}$. The predicted
deviations of the evaporation energy from  the liquid drop behaviour
are rather small (less than $0.1$ K).  Unfortunately this value is
smaller than the temperature of droplets in available experimental beams
(about $0.4$ K).

\subsection{Layering and prewetting transitions in films.}
\label{sec:films}

	In order to find the equilibrium state of liquid helium on a solid
substrate, we add the external helium-substrate potential
$V_{sub}(\rr)$ to the mean field $U$ in the Hartree equation
(\ref{eq:hartree}).  We assume the substrate to be flat, avoiding the
problem of possible corrugations.  This approximation is certainly  valid
for weak-binding  substrates, such as the alkali metals,
for which the cloud of delocalized  electrons is expected to smooth out
the potential along the substrate plane. In  this case the Hartree
equation is again one dimensional.  The substrate-helium potential is
taken here as a 9-3 potential of the form
\begin{equation}
        V_{sub}(z) = {4 C_3^3 \over 27 D^2 z^9}-
        {C_3 \over z^3} \ \ \ ,
\label{eq:vsub}
\end{equation}
where $C_3$ is the Hamaker constant and $D$ the well depth of the
potential. The values taken from \cite{Vid91} for various substrates
are gathered in Table \ref{tableI}.

	The energetics of helium films on various surfaces indicate that two
types of phase transition in film growth  can take place, depending on the
strength of the substrate potential. On  strong and medium binding surfaces,
which create large local pressures  (one or two layers may become solid) the
growth of the first liquid  layers proceeds via {\it layering transitions},
first described for  helium on graphite in Ref.~\cite{Clem93}. When considering
weak  binding substrates, however, these layering transitions are no longer
present. One enters a new regime of film adsorption where {\it prewetting
transitions} take place at $T=0$, as analyzed in details in
Ref.~\cite{Cheng1}. By convention, the term layering  transition is reserved
usually to the case of a first order
transition  in film thickness involving {\it one
layer} only. The occurence of such  transition is not linked to the question of
wetting; it is essentially  related to the nature of the quasi-2D system. To
the contrary, a prewetting  transition involves a jump in film thickness which
can take {\it any  value}. The notion is intimately connected to that of
wetting, since a  prewetting transition is the continuation of the wetting
transition off  coexistence.

	The various cases are best illustrated by considering, for a given
substrate, the dependence of the chemical potential $\mu$ on coverage, as shown
in Fig.~\ref{fig:mu_film}. A negative slope ($d\mu/dN < 0$) indicates an
unstable range of film thickness, and the transitions are determined by Maxwell
constructions. The figure shows the results for a $H_2$ substrate and for the
alkali metals. For $H_2$, we have used a parametrization of the
helium-substrate interaction proposed in Ref.~\cite{Him87}:
\begin{equation}
        V_{H_2}(z) = {900000 \over z^9}-{15000 \over z^5}
     			-{435 \over z^3} \ \ \ ,
\label{eq:vh2}
\end{equation}
where $z$ is in \AA~and $V_{H_2}$ in K. This form is fitted to the results of
Pierre {\it et al.} \cite{Pier85} and gives a well depth of 33~K. The binding
energy of one \hef atom is 16.4~K, in good agreement with the experimental
determination of Paine and Siedel \cite{Pai92}, which is 16$\pm$2~K. For this
potential, two layering transitions occur. The mechanism by which these
transitions are produced can be summarized as follows: on a strong substrate,
liquid helium forms well defined layers which are  approximately independent
quasi-2D systems. As 2D helium is a liquid  (Monte-Carlo simulations
\cite{Whi88} give a binding energy of 0.8 K at an  equilibrium density of 0.043
\AA$^{-2}$) the formation of each of the  first layers exhibits a quasi-2D
condensation. It is important to  realize that the two aspects are necessary :
on one hand the layering  transitions would not occur if quasi-2D $^4$He were a
gas; on the other hand we  shall see below that they also disappear if the
substrate is not strong  enough to produce a sufficient layering of the fluid,
as is the case  with the alkalis. Fig.~\ref{fig:prof_h2} represents the growth
of a helium film on $H_2$ characterized by two regions of instability or
metastability.

	On sees from Fig.~\ref{fig:mu_film} that the Orsay-Paris functional
produces rather smooth curves and in particular misses the layering transitions
besides the first one. This fact is due to two deficiencies of the model,
related to the mechanism described above, namely i) the peak of the
density-density response function $\chi(q)$ is underestimated, as we have seen,
by a factor of almost 2 and ii) the binding energy of the quasi-2D system is
too small. Both defficiencies are corrected with the new functional.

	Let us now turn to the results for the alkalis. Although  $\mu(N)$ has
still some structure, Mg  appears as a limiting case where the layering
transitions tend to disappear, except for the first one. Interestingly, one
sees that with  decreasing strength of substrate potential, it is this first
transition  which becomes larger and larger in amplitude and thus turns into
the  prewetting transition. The physics here is no longer that of the quasi-2D
system, but that of wetting and prewetting : for a given substrate, the
thinnest stable film is such that the energy cost of forming two interfaces
-one with the substrate, one free surface- is compensated by the energy gain of
placing the fluid in the attractive potential of the wall. The limit between
wetted and nonwetted substrates is obtained when  stability is obtained only
for an infinitely thick film. Notice that the  predictions of the Orsay-Paris
functional and of the present one become similar for these weak binding
surfaces. Using the original values of  the substrate potential parameters, one
still finds that the three  alkalis Cs, Rb and K are not wetted. The new
functional  slightly favors wetting with respect to the Orsay-Paris one. For
example, the contact angle calculated for Cs is reduced by 3 degrees; also, on
a Na substrate, the prewetting jump is reduced to 3.3 layers, compared to  5.2,
and metastable films are found to exist down to 1.2 layers.

	Layering growth can be also seen in the adsorption isotherms. In
Fig.~\ref{fig:adsorp} is plotted the isotherm $T$ = 0.639 K for helium on
graphite. For comparison with experimental data \cite{Zim92} we use the ideal
gas formula
\begin{equation}
{ P \over  P_0} = \exp \left[ { \mu-\mu_0 \over  k_B T} \right] \ \ \ ,
\label{eq:isotherme}
\end{equation}
where $\mu_0=-7.15$ K is the bulk chemical potential, $P_0$ the
saturating vapor pressure at $T$. The graphite substrate is a test for the
model in a highly inhomogeneous situation. The two first layers are known to
be solid. To first approximation, the effect of localization of the atoms in
the plane parallel to the  substrate can be ignored. Hence, we have treated the
two solid layers as the liquid. The experimental data show a clear staircase
structure, associated with the progressive filling of layers. The results of
the present  functional (solid line) exhibit a similar pattern, with steps of
amplitude and phase close to the experimental ones.

\section{Dynamics}
\label{sec:dynamics}

In Sect.~\ref{sec:statics} we wrote the Hartree equation (\ref{eq:hartree})
for the ground state of the fluid, which corresponds to the minimization
of the energy $E$ in Eq.~(\ref{eq:e}) with respect to the density.  This
formalism is generalized to the study of dynamical properties using the
least action principle:
\begin{equation}
\delta \int_{t_1}^{t_2} dt \int \! d\rr \left[ {\cal H} [\Psi^*, \Psi]
-\mu \Psi^* \Psi - \Psi^* i \hbar {\partial \Psi \over \partial t }
\right] \ =\ 0
\ \ \ .
\label{eq:leastaction}
\end{equation}
The equations of motion for the excited states of the fluid can be
derived by making variations with respect to $\Psi$ or $\Psi^*$. One
finds  a Schr\"odinger-like equation of the form
\begin{equation}
	(H - \mu )  \Psi = i \hbar {\partial \over \partial t } \Psi \ \ \ ,
\label{eq:schro}
\end{equation}
where $H = \delta E/ \delta \Psi^*$ is an effective
Hamiltonian.  We linearize the equation by writing
\begin{equation}
	\Psi ({\bf r},t) = \Psi_0({\bf r}) + \delta \Psi({\bf r},t)
\label{eq:deltapsi}
\end{equation}
where $\Psi_0({\bf r})$ refers to the ground state. The Hamiltonian
$H$ then takes the form
\begin{equation}
	H = H_0 + \delta H \ \ \ .
\label{eq:deltah}
\end{equation}
The static Hamiltonian
\begin{equation}
	H_0  = - {\hbar^2 \over 2m} \nabla^2 + U[\rho,\rr]
\label{eq:htilde0}
\end{equation}
which appeared already in Eq.~(\ref{eq:hartree})  determines the
equilibrium state $\Psi_0({\bf r}) = \sqrt{\rho(\rr)}$. The term
$\delta H$ is linear in $\delta \Psi$ and accounts
for changes in the Hamiltonian induced by the collective
motion of the system. Since $H$ depends explicitly
on the wave function $\Psi$, the Schr\"odinger
equation (\ref{eq:schro}) has to be solved using a self-consistent
procedure,  even in the linear limit considered in the present work.

The formalism here described corresponds to a time dependent density
functional (TDDF) theory which, in the linear limit (\ref{eq:deltapsi}),
coincides with the Random  Phase  Approximation (RPA) for a Bose
system. This theory, which  is basically a mean field theory
coupling one-particle/one-hole configurations,  is suitable for
describing collective
(one-phonon) states, but not multiphonons excitations.

A completely equivalent formulation of the equations of motion can be
obtained by using the canonically conjugate variables $\rho$ and  $S$,
defined in Eq.~(\ref{eq:psi}). The least action principle takes the form
\begin{equation}
\delta \int_{t_1}^{t_2} dt \int \! d\rr \left[
{\cal H}_0[\rho] + {\cal H}_v[\rho,{\bf v}]
-\mu \rho + \rho {\partial S \over \partial t }  \right]
\ =\ 0  \ \ \ ,
\label{eq:leastaction2}
\end{equation}
where we have separated the velocity dependent part of the
functional from the velocity independent one.   After variations
with respect to  $\rho$ and $S$, one finds two coupled equations of the
form
\begin{eqnarray}
	\frac{\partial \rho}{\partial t} &+& \frac{1}{m}
	\nabla_\rr \ \frac{\delta}{\delta {\bf v}} \int d\rr \left\{
	{\cal H}_v[\rho, {\bf v}] \ \right\} = 0
\label{eq:cont}\\
	\frac{\partial S}{\partial t} &+& \frac{\delta}{\delta \rho} \ \int
	d\rr
	\left\{  {\cal H}_0[\rho] + {\cal H}_v[\rho, {\bf v}] - \mu \rho
	\right\} = 0 \ \ \ ,
\label{eq:euler}
\end{eqnarray}
\noindent which can be viewed as a generalization of the equation of
continuity  and the Euler equation. In the hydrodynamic limit,
${\cal H}_v$ is given by the usual kinetic energy term
\begin{equation}
{\cal H}_v[\rho,{\bf v}] = {m \rho \over 2} |{\bf v}|^2 \ \ \ ,
\label{eq:mv2}
\end{equation}
so that (\ref{eq:cont}) leads to the usual conserved current
\begin{equation}
	{\bf J}_0(\rr) = \rho(\rr) {\bf v}(\rr)
	= \frac{\rho(\rr)}{m} \nabla S(\rr) \ \ \  .
\end{equation}

The present formalism, with ${\cal H}_v$ given by (\ref{eq:mv2}) and
with different choices for ${\cal H}_0$, has been already applied
to helium droplets  \cite{Cas90,Bar94} and films \cite{Pri94}.
However, there are two major shortcomings which make the
results of those  calculations not satisfactory from a quantitative
point of view.  First, the
static response  function is not well reproduced in the roton region.
Second, the theory does not account for backflow effects.

	To understand this point better, let us discuss the results of
the TDDF in bulk. In this case the density is
constant and $\delta \Psi$ can be expanded in plane waves, corresponding
to the  propagation of phonon-roton excitations. In the absence of backflow
effects, the only  dependence of the energy on the velocity
field comes from the hydrodynamic limit of ${\cal H}_v$
(see Eq.~(\ref{eq:mv2}) and the resulting dispersion relation for
the phonon-roton mode takes the form
\begin{equation}
\hbar \omega (q) \  = \ \left( { \hbar^2 q^2  \over m |\chi(q)| }
			\right)^{1/2}
\ \ \ .
\label{eq:omega}
\end{equation}
Result (\ref{eq:omega}) can be also written in terms of the moments
(\ref{eq:moments}) of dynamic structure function $S(q,\omega)$. Indeed,
one has
\begin{equation}
\hbar \omega (q) \ = \ \left( {m_1(q) \over m_{-1} (q) } \right)^{1/2}
\ \ \ ,
\label{eq:m1overm-1}
\end{equation}
where the energy-weighted moment
\begin{equation}
m_1(q) \  =\  \int_0^\infty d\omega \ S(q,\omega) \, \hbar \omega \ = \
{\hbar^2 q^2 \over 2 m}
\label{eq:fsumrule}
\end{equation}
coincides with the well known f-sum rule, while the inverse
energy-weighted moment $m_{-1}$ is related to the static response
function  by  the compressibility sum rule (\ref{eq:chi}).

	Density functionals having the form (\ref{eq:mv2})  for  ${\cal H}_v$
exactly fullfil  the f-sum rule. Thus, differences in the predictions for the
dispersion $\omega(q)$ come only from the quantity  $\chi(q)$, which is fixed
by the static part  ${\cal H}_0[\rho]$ of the functional.  In
Fig.~\ref{fig:bulk} we show the results for the phonon-roton dispersion
obtained with the Orsay-Paris functional (dashed line) and functional
(\ref{eq:h0}) (dot-dashed line). Both curves overestimate significantly
the experimental phonon-roton energy (points). The difference between the
predictions of the two  functionals  is clearly understood by looking at
Fig.~\ref{fig:chi} and  Eq.~(\ref{eq:omega}): starting from the  hamiltonian
density of Eq.~(\ref{eq:h0}), one obtains  the full peak of $\chi(q)$ at the
roton wavelength, thus predicting a lower roton energy.  The remaining gap
between  theory and experiment is mainly due to the role of multiphonon
excitations in the f-sum rule. In fact, the analysis of the  spectra of neutron
scattering experiments \cite{Cow71} shows that the collective mode gives only a
fraction ($\simeq 1/3$) to the f-sum rule in the roton region, the remaining
part being  exhausted by high energy  multiphonon excitations. On the contrary,
due to the $\omega^{-1}$ factor in the integrand, the collective mode almost
exhausts the compressibility  sum rule for wavelengths up to about $2.2$
\AA$^{-1}$. This means that in order to have an accurate prediction for
$\omega(q)$, only the  single mode contribution to the f-sum rule should be
included  in the moment $m_1(q)$ entering  Eq.~(\ref{eq:m1overm-1}). Therefore,
one concludes that, once the static response function $\chi(q)$ is properly
accounted for, the dispersion law (\ref{eq:omega}) given by TDDF theory
provides an upper bound to the exact energy of the phonon-roton mode.

It is instructive to compare the above predictions with the results
of the so called Feynman approximation for collective excitations
\cite{Fey54}, often  used in dynamic calculations for non uniform
$^4$He states \cite{Kro92,Wha90,Chi92,Ger92}.  In the bulk the
Feynman dispersion law takes the form
\begin{equation}
\hbar \omega_F (q) \ = \ {m_1(q) \over m_0 (q) }
\  = \ {\hbar^2 q^2 \over 2 m S_q}
\ \ \ ,
\label{eq:m1overm0}
\end{equation}
where the static structure factor is related to the non energy-weighted
moment of $S(q,\omega)$ through the equation
\begin{equation}
S (q)  = m_0(q) = \int\! d\omega S(q,\omega) \ \ \ .
\label{eq:sq}
\end{equation}
General properties of the moments $m_n$ permit to prove the following
inequality:
\begin{equation}
\sqrt{ {m_1 \over m_{-1}} } \le {m_1 \over m_0 }
\label{eq:inequality}
\end{equation}
holding at zero temperature. This implies that the dispersion law
given by Eqs.~(\ref{eq:omega},\ref{eq:m1overm-1})  provides
an upper bound closer to the exact dispersion law with respect to the
Feynman approximation. This is explicitly shown in Fig.~\ref{fig:bulk}.

Concerning the static structure factor $S(q)$, one should note that it
can not be properly accounted for by the TDDF theory developed above.
The reason is that, in the bulk liquid, this theory represents a single
mode approximation with the dynamic structure function in the form
\begin{equation}
S^{\scriptstyle DF}(q,\omega) = {\hbar q^2 \over 2m \omega_{\scriptstyle DF}}
\delta (\omega -\omega_{\scriptstyle DF})
\label{eq:sdf}
\end{equation}
and the dispersion law $\omega_{\scriptstyle DF}$ given by
Eq.~(\ref{eq:m1overm-1}). While Eq.~(\ref{eq:sdf}) reproduces exactly the
sum rules $m_1$ and $m_{-1}$, it yields the approximate expression
\begin{equation}
S^{\scriptstyle DF} (q) \ = \ \sqrt{ m_1(q) m_{-1}(q) }\  = \
\sqrt{ {\hbar^2 q^2 \over 4 m} |\chi(q)| }
\label{eq:sqdf}
\end{equation}
for the static structure factor. Actually result (\ref{eq:sqdf})
provides an upper bound to the exact value of $S(q)$. The difference
is again due to the role of multiphonon excitations.

	One of the purposes of this work is to make the TDDF
theory more quantitative. The idea is to realize that the expression of
${\cal H}_v$ given by Eq.~(\ref{eq:mv2}) comes from a many-body wave
function of the form
\begin{equation}
	\Psi_N(\rr_1,... \rr_N) = \exp \left [i \sum_i s(\rr_i) \right]
       	|\Psi_N(\rr_1,... \rr_N)| \ \ \ ,
\label{psi:gas}
\end{equation}
where $s$ is real. Clearly, this wave function does not take into account short
range phase correlation, and describes correctly the superfluid motion in the
hydrodynamic limit only. The exact wave function should be expressed
with the more general phase:
\begin{equation}
	s({\bf r_1,r_2, ...r_N}) = \sum_i s_1(\rr_i)
	       	+ \sum_{i<j} s_2(\rr_i,\rr_j) +...
	      	+ \sum_{i<j<k...} s_N(\rr_i,\rr_j,\rr_k,...) \ \ \ .
\end{equation}
The average kinetic energy obtained from this wave function has non-local
contributions coming from $s_2,... s_N$. Now, if we make the assumption that
those terms can be expressed only with the two canonically conjugate variables
$\rho$ and $S$, then we are led to add a non-local velocity dependent term to
${\cal H}_v[\rho, {\bf v}]$. Indeed, this procedure has been proposed a long
time ago by Thouless \cite{Tho69} in the study of the flow of a dense
superfluid, but in his article, helium was treated as an incompressible liquid.
In the present work, we incorpore the suggestion of Thouless by
introducing the most general quadratic form
\begin{eqnarray}
      {\cal H}_v = {m \over 2} \rho(\rr)
                     |{\bf v}({\bf r})|^2
     + \int \!  d{\bf r}_1 d{\bf r}_2 d{\bf r}_3 \
     		[{\bf v}({\bf r})-{\bf v}({\bf r}_1) ]
	\underline{G} (\rho;{\bf r}, {\bf r}_1, {\bf r}_2, {\bf r}_3)
     	[{\bf v}({\bf r}_2)-{\bf v}({\bf r}_3) ] \ \ \ ,
\label{eq:thouless}
\end{eqnarray}
where we have explicitly exhibited Galilean invariance. Generally  speaking,
$\underline{G}$ is a tensor which takes non zero values on scales $|{\bf
r}_i-{\bf r}_j|$ of the order of the inter-particle distance.  So far no
microscopic derivation has been found for  this functional form. Our
phenomenological approach consists in keeping only diagonal terms as follows:
\begin{equation}  {\cal H}_v = {m \over 2} \rho(\rr) |{\bf v}({\bf r})|^2
             - {m \over 4} \int \! d\rp \ V_J (|\rr-\rp|) \ \rho(\rr)
	     \rho(\rp) \  \left[ {\bf v}(\rr) -{\bf v}(\rp) \right]^2 \ \ \ ,
\label{eq:hv}
\end{equation}
and fixing the effective current-current interaction $V_J$ to reproduce
known properties in bulk.  The new term plays the role of a  non local
kinetic energy. Now, the conserved current is no longer ${\bf J}_0$;
rather, Eq.~(\ref{eq:cont}) leads to the current
\begin{eqnarray}
	&{\bf J}&(\rr) = {\bf J}_0(\rr) + {\bf J}_B(\rr), \\
        &{\bf J}&_B(\rr) = \rho(\rr) \int \rho(\rp) \left[
	{\bf v}(\rr)-{\bf v}(\rp) \right] V_J(|\rr-\rp|) d\rp
\end{eqnarray}
Physically, ${\bf J}_B(\rr)$ acts as a backflow which depends on the
velocity and the
density in the vicinity of point $\rr$. As expected, its contribution vanishes
when many-body phenomena are not present ($\rho \to 0$).

	This form of ${\cal H}_v$ lowers the value of the energy weighted
moment  $m_1(q)$ predicted by the density functional theory. In bulk one
now finds
\begin{equation}
m_1(q) = \frac{\hbar^2 q^2}{2 m} \left\{ 1 - \rho
	\left[ \hat{V}_J(0) - \hat{V}_J(q) \right] \right\},
\label{eq:m1vj}
\end{equation}
where $\hat{V}_J (q)$ is the Fourier transform of the current-current
interaction $V_J(r)$. Notice that the expression for the static
response function $\chi(q)$ does not change, since it is entirely fixed by
${\cal H}_0$. The phonon-roton dispersion in bulk is still given by the
ratio (\ref{eq:m1overm-1}), so that the dispersion law is given by
\begin{equation}
\left[ \hbar \omega (q) \right]^2 \  = \  \frac{\hbar^2 q^2 \rho}
 {m |\chi(q)| }
 \left\{ 1 - \rho \left[ \hat{V}_J(0) - \hat{V}_J(q) \right] \right\}
\ \ \ .
\label{eq:omegavj}
\end{equation}
This relation can be used to fix $V_J(r)$ in order to reproduce
phenomenologically the experimental phonon-roton dispersion.
We have chosen the simple parametrization
\begin{equation}
 V_J (r) = (\gamma_{11} + \gamma_{12} r^2) \exp(-\alpha_1 r^2)
         + (\gamma_{21} + \gamma_{22} r^2) \exp(-\alpha_2 r^2) \ \ \ ,
\label{eq:vj}
\end{equation}
where the parameters are given in Table \ref{tableII}. The corresponding
dispersion relation is shown in Fig.~\ref{fig:bulk} (solid line).  The
roton minimum  is at $q_0=1.92$ \AA$^{-1}$ and the roton energy is $\Delta
= 8.7$ K.   The  pressure  dependence of the dispersion relation turns out
to be also  well reproduced. At $P=24$ atm, for instance, the roton
minimum is displaced  at $q=2.01$ \AA$^{-1}$ and  the roton energy is
$\Delta=7.4$ K, close to the  experimental values  $q=2.05$ \AA$^{-1}$
and $\Delta=7.3$ K \cite{Sti90}.

	The energy weighted moment $m_1(q)$ is shown in Fig.~\ref{fig:m1},
where the density functional result (\ref{eq:m1vj}) is compared with the
experimental data for the collective contribution to the f-sum rule
\cite{Cow71}. The fact that the new functional no longer  satisfies the
f-sum rule (\ref{eq:fsumrule}) points out in a clear way that  the TDDF
theory does not account for multiphonon excitations. The new current-current
term in the functional (\ref{eq:hv}) changes also the expression for
the static structure factor,  which is still given by $S^{\scriptstyle DF}(q)
= \sqrt{ m_1(q) m_{-1}(q)}$ as in Eq.~(\ref{eq:sqdf}), but with the new
$m_1(q)$ moment (\ref{eq:m1vj}). The resulting $S^{\scriptstyle DF}(q)$ is
no more an upper bound to the exact $S(q)$; conversely it turns out to be
close to the experimental one-phonon contribution to the total $S(q)$.
A similar separation between collective and  multiphonon
excitations, in the context of linear response  theory, has been developed
by Pines and collaborators \cite{Pin83}. Finally we note that the
dispersion law (\ref{eq:omegavj}) is not  affected by the new term in the
$q \to 0$ hydrodynamic regime,  where it gives the usual sound velocity
$c^2=- [m\chi(0)]^{-1}$. This is  an important feature ensured by
Galilean invariance.

 In conclusion, the complete Orsay-Trento functional has the form
\begin{equation}
E \ = \ \int \! d\rr \ \left\{ {\cal H}_0 [\rho]  +
{\cal H}_v [\rho,{\bf v}] \right\}
 \ \ \ ,
\label{orsay-trento}
\end{equation}
where ${\cal H}_0$ and ${\cal H}_v$ are given in  Eqs.~(\ref{eq:h0}) and
(\ref{eq:hv}) respectively. The term ${\cal H}_v$ vanishes in the
ground state calculations of the previous sections. On the contrary, it
is crucial in the calculation of the dynamics. Both the experimental
static response function and the phonon-roton dispersion in bulk are
taken as input to parametrize the functional. The theory can then be
applied to study the excited states of inhomogeneous systems. This is
the purpose of the next section.

\section{Excited states of the free surface and of films}
\label{sec:excited}

\subsection{Equations of motion.}
\label{planar}

	A detailed study of excitations using the Orsay-Paris functional
can be found in Ref.~\cite{Pri94}. The same method of resolution is
followed here. The systems under consideration have translational
invariance parallel to the $x-y$ plane, hence the ground state wave
function depends only on one coordinate, which is taken as the
$z$-coordinate orthogonal to the free surface or to the plane of a film.
The excited state wave functions can be expanded in  plane waves as
follows :
\begin{equation}
\Psi(\rpar,z,t) = \Psi_0(z) + \sum_{\bk,b} \frac{1}{\sqrt{AL}}
\{ \alpha_{\bk,b}\Phi_{\bk,b}^{1}(z)
	e^{-i(\omega_{k,b} t-\bk\cdot\rpar)} +
\alpha_{\bk,b}^{*}\Phi_{\bk,b}^{2}(z)
	e^{i(\omega_{k,b} t-\bk\cdot\rpar)}\} \ \ \ ,
\label{eq:psixzt}
\end{equation}
\noindent where $\rpar$ denotes from now on a two-dimensional vector
parallel to the surface of the liquid. The quantities $\alpha_{\bk,b}$ are
small amplitudes, the factor $1/\sqrt{AL}$ is a normalization constant
($A$ is the area of the sample, $L$ is an arbitrary length) and $b$ labels
the various branches of excited states. The functions $\Phi_1 (z)$ and
$\Phi_2(z)$, without any loss of generality, can be chosen real. This
formalism has been already used in Refs.~\cite{Pri94,Ji86}.

It is convenient to introduce the quantities
\begin{equation}
	\Phi_{k,b}^{\pm}=\Phi_{k,b}^{1} \pm \Phi_{k,b}^{2*} \ \ \ .
\label{phi}
\end{equation}
Then the  phase $S$ can be written as
\begin{equation}
S(\rpar,z,t) = - \frac{i\hbar}{2\sqrt{AL}} \sum_{\bk,b}
	\frac{1}{\Psi_0(z)} \{ \alpha_{\bk,b} \Phi_{{\bf k},b}^-(z)
      	e^{-i(\omega_{k,b}t-\bk\cdot\rpar)}
        \} + c.c.
\label{phase}
\end{equation}
and the density takes the form
\begin{equation}
 \rho(\rpar,z,t) = \rho_{0}(z)
			+ \delta\rho(\rpar,z,t)
\end{equation}
where $\rho_{0} = |\Psi_0|^2$ is the ground state density, while
\begin {equation}
	 \delta\rho(\rpar,z,t)
 = \frac{1}{\sqrt{AL}}\sum_{\bk,b} \Psi_0(z) \{\alpha_{\bk,b}
\Phi_{\bk,b}^+(z) 	e^{-i(\omega_{k,b} t-\bk\cdot\rpar)} \} + c.c
\label{deltaro}
\end{equation}
The equations of motion (\ref{eq:cont}) and (\ref{eq:euler})
assume the form
\begin{equation}
\left\{ 	\begin{array}{l}
	{\cal D}_k\Phi_{k,b}^+ +
	{\cal W} \big\{ \Phi_{k,b}^+ \big\}
	= \hbar \omega_{k,b} \Phi_{k,b}^{-}
\\
	{\cal D}_{k}\Phi_{k,b}^{-} +
	{\cal U} \big\{ \Phi_{k,b}^- \big\}
	= \hbar\omega_{k,b} \Phi_{k,b}^{+}
      		\end{array}
\right.
\label{eq:system}
\end{equation}
The eigenvalues $\hbar\omega_{k,b}$ satisfying Eq.~(\ref{eq:system}) are the
energies of the collective modes of the system. The symbol ${\cal D}_k$
denotes a  differential operator of second order, while ${\cal W}$ and
${\cal U}$ are integral operators. They are defined by
\begin{eqnarray}
	& & {\cal D}_k = -\frac{\hbar^{2}}{2m} \frac{d^{2}}{dz^{2}}
+ \frac{\hbar^{2} k^{2}}{2m} + U[\rho_0,z] + V_{sub}(z)   - \mu
\label{dk}\\
	& & {\cal W} \big\{ f \big\} = 2 \Psi_0(z) \int d\rp
 \Psi_0(z') f(z') \left[\frac{\delta^2 E_c}{\delta\rho(\rr)
\delta\rho(\rp)} \right] e^{i\bk\cdot(\rpar-\rpar')}
\label{noy1}\\
	& & {\cal U} \big\{ f \big\} = \frac{\dis 1}{2 \Psi_0(z)} \int d\rp
 \frac{\dis f(z')}{\Psi_0(z')} \left[\frac{\delta^2 E_v}{\delta S(\rr)
\delta S(\rp)} \right] e^{i\bk\cdot(\rpar-\rpar')}
\label{noy2} \ \ \ ,
\end{eqnarray}
where $E_v$ is a functional containing only the non-local part of ${\cal H}_v$.
A simple way of solving Eq.~(\ref{eq:system}) is to expand it on the
basis of eigenstates of the static one-body Hamiltonian. This leads  to a
matrix equation that is easily solved numerically by a direct diagonalization.

A quantum hydrodynamical formalism is also obtained by making the
substitution $\{-i \alpha_{\bk,b} \} \rightarrow a_{\bk,b}$, where
$a_{\bk,b}$ is the operator of creation of one phonon which statisfies the
Bose commutation rules
\begin{equation}
	[a_{\bk,b},a_{\bk',b'}^{\dag}]
	= \delta_{\bk,\bk'} \delta_{b,b'}
	\ \ \ \ \ \ \
	[a_{\bk,b},a_{\bk',b'}] = 0 \ \ \ .
\nonumber
\end{equation}
{}From the completeness relation
\begin{equation}
	\sum_b \Phi_{k,b}^-(z) \Phi_{k,b}^+(z') = L \delta(z-z')\ \ \ .
\label{compl}
\end{equation}
one   recovers the equalities
\begin{equation}
	[\rho(\rr),S(\rp)] = i\hbar\delta(\rr-\rp)
	\ \ \ \ \ \ \
	[\rho(\rr),\rho(\rp)] = [S(\rr),S(\rp)]=0
	\ \ \ ,
\label{commut}
\end{equation}
Then, after a quadratic expansion of the total energy and using
the properties of Eq.~(\ref{eq:system}), one finds  the Hamiltonian
\begin{equation}
	H^{(2)} = E_{0} + \sum_{\bk,b}
		\hbar\omega_{\bk,b} a_{\bk,b}^{\dag} a_{\bk,b} \ \ \ ,
\end{equation}
\noindent where $E_{0}$ is the energy of the ground state. This
diagonalization is obtained provided the $\Phi_{k,b}^{\pm}$'s are
normalized according  to
\begin{equation}
	\int dz \, \Phi_{k,b}^-(z) \Phi_{k,b'}^+(z) = L \delta_{b,b'} \ \ \ ,
\label{ortho}
\end{equation}
\noindent where the orthogonality appears as a consequence of
(\ref{eq:system}). Higher order expansions of $H$ in terms of  density and
phase fluctuations would give  rise to interactions between
quasi-particles.

	The transition density associated  with the  solution
(\ref{deltaro}) is given by (the vector ${\bf q}$ denotes $(\bk,q_z)$)
\begin{equation}
	\left(\rho^{\dag}_{\bf q}\right)_{\bk,b}
	= -i \sqrt{\frac{\dis A}{\dis L}}
	\int \Psi_0 (z) \Phi_{k,b}^+(z) \exp(iq_z  z) dz
\label{eq:drho}
\end{equation}
and the dynamic structure function can be evaluated by means of
the definition
\begin{equation}
	S(k,q_z,\omega) =  \sum_b \left|  \left(\rho^{\dag}_{\bf q}
\right)_{\bk,b} \right|^2  \delta(\omega - \omega_{\bk,b}) \ \ \ .
\label{eq:sqomega2}
\end{equation}
In order to compare the theoretical strength with the experimental
results, we have introduced a width of the order of the experimental
resolution, by substituting the $\delta$-function of Eq.~(\ref{eq:sqomega2})
with a normalized gaussian,  as  done in  Ref.~\cite{Cle94}.

\subsection{Dynamics of the free surface.}

We have done calculations in slab geometry, i.e., liquid between two
parallel surfaces at a distance $L$ much larger than the surface thickness
(typically 50 to 100 \AA). The slab geometry is a good approximation
to  a semi-infinite medium  for wave vectors larger than $1/L$.  The
dispersion relation of the various modes, extrapolated to $L \to \infty$,
are shown in  Fig.~\ref{fig:freesurface}. The lowest branch corresponds
to a wave function $\Phi^+(z)$ localized in the surface region.
In the long  wavelength limit, it coincides with the hydrodynamical surface
wave called ripplon, whose dispersion is
\begin{equation}
\omega^2 (k)  = {\sigma k^3 \over m \rho_\infty } \ \ \ ,
\label{ripplon}
\end{equation}
where $\sigma$ and $\rho_\infty$ are the surface tension and the bulk
density respectively. For wave vectors of the order of 0.5 \AA$^{-1}$, the
dispersion relation starts deviating from the hydrodynamical limit,
until its curvature eventually changes sign, due to a coupling with bulk modes.
Our curve reaches the value of the roton energy $\Delta=8.7$ K at about
$k=1.15$ \AA$^{-1}$. A first experimental evidence of a similar deviation
from the hydrodynamic law came from measurements of the surface
entropy \cite{Edw74}. More recently the dispersion of surface
modes has been measured in neutron scattering
experiments on helium films \cite{Lau90}. The experimental data are
shown in  Fig.~\ref{fig:freesurface} as squares, the error bars
accounting for the spreading of the data for different coverages
($3.5$ to $5$ layers of helium on graphite).  The agreement with the
calculated values is satisfactory.

Below $\Delta$ the surface modes are undamped, while above $\Delta$
they couple with the continuum of bulk modes (rotons with negative and
positive group velocity) propagating at different angles ($q_z \neq 0$).
This results in a spreading of the strength associated with the
surface modes. Actually the spreading predicted by our theory is
small. In Fig.~\ref{fig:sqomegafree} we show the dynamic structure
function for scattering at grazing angle ($q_z=0$) on a slab
of thickness $L=50$ \AA.  The strength
of the lowest surface mode is well localized not only below but also
above $\Delta$, even though it is partially distributed among
bulk modes coupled to ripplons. The position of the peak of the ripplon
mode above $\Delta$ is shown also in Fig.~\ref{fig:freesurface} as a
dashed line.

We obtain also a second branch of surface excitations, lying in between the
ripplon and  the phonon-roton modes. It is visible in
Fig.~\ref{fig:sqomegafree} as a small bump, which is close to the phonon peak
for $k \simeq 0.3$ \AA$^{-1}$ and stays almost parallel to the ripplon
dispersion for larger $k$. The relative strength of this mode is larger
in thin films; in that case, the experiments seem also to
support  the existence of such a surface mode.

	A more detailed analysis of the results of the present density
functional theory for the excitations of the free surface is given in
Ref.~\cite{Las94}; in that work, a general discussion of the
mechanism of hybridization between ripplons and rotons is presented, and
some properties connected with reflection and evaporation of bulk excitations
impinging on the surface are also discussed.

\subsection{Dynamics of films.}

	In order to illustrate the results obtained with the
present model, we have chosen the case of a H$_2$ substrate, since it
has been the subject of both experimental and theoretical investigations
\cite{Cheng3,Shi91}. Besides finite size effects, the interesting
features to be expected are linked to the layering of the liquid near
the substrate. Submonolayer superfluidity has been observed on H$_2$
\cite{Shi91}, indicating that helium remains liquid close to the
substrate. However the well-depth of the helium-hydrogen potential is
rather large (33~K), which produces a well defined layering of the
liquid.

	In a thin film, the long wavelength limit of the lowest
surface mode is no longer a ripplon, for which the restoring force
originates from the surface  tension, but rather a third sound
excitation. The restoring force is here given by the substrate
potential. The third sound speed $c_{3s}$ is obtained as  the hydrodynamic
limit ($\bk
\rightarrow 0$) of Eq.~(\ref{eq:system}):
\begin{equation}
m c_{3s}^2 = N \frac{d \mu}{d N},
\end{equation}
where $N$ is the coverage of the film. Indeed, one can verify that for
small momentum, $\Phi^- \sim \Psi_0$, so that the contribution due to
${\cal U}$ vanishes. With increasing film thickness, since the chemical
potential varies as $1/N^3$, so does $c_{3s}^2$.

	The results for $c_{3s}$ (Fig.~\ref{fig:c3_H2}), show strong
oscillations as a function of coverage. This is due to the  layered structure
of the film, which is reflected into the coverage  dependence of the chemical
potential (see Sect.~\ref{sec:films} above). Films  for which a uniform
coverage is unstable have a constant chemical  potential characteristic of a
layering transition, the range of which is determined by a Maxwell construction
as discussed above. This causes the third sound velocity to drop dramatically
to zero. The  structure is more marked than with the Orsay-Paris functional
\cite{Cheng3}, which missed the layering transitions besides the first  one
(see Sect.~\ref{sec:films}).  Experimentally, there is up to now no indication
of the layering  transitions. The third sound velocity does show oscillations
\cite{Shi91},  however less marked than calculated here, although the
measurements were  done at a temperature of 0.18 K, which is expected to be
lower than the critical temperature of the layering transitions. It is also
possible that surface inhomogeneities of the H$_2$ surface are able to smooth
out the  dips seen in the calculations of $c_{3s}$ for a perfectly flat
surface.

	Two typical spectra are shown in Fig.~\ref{fig:spec_h2} for two
values of helium coverage ($N=0.24$ \AA$^{-2}$  and $N=0.48$ \AA$^{-2}$)
on H$_2$ substrate. For wave vectors  in the range $1.5$ to $2$ \AA$^{-1}$,
the low lying excited states are modes confined in the first layers of the
fluid close to the substrate. Their nature has  been discussed in details
in Refs. \cite{Pri94,Cle94}. Their formation reflects the fact that the
substrate-liquid interface is nearly solidified, so that low energy
excitations appear, trapped in each of the first layers, with a momentum
corresponding to the interparticle distance. The minimum 2D roton energy
was found to be about $12$ K with the Orsay-Paris functional. It reduces
to $5.6$ K with the present one. The combined effect of the evolution of the
third sound excitation and of the 2D roton, with completion of the first layer,
may lead to unexpected behaviour of the heat capacity of submonolayer films
\cite{Pri94}.

	In Fig.~\ref{fig:strenght_h2}, we show the dynamic structure function
associated with the spectra of Fig.~\ref{fig:spec_h2}. The strength
distribution of the thinner film is more fragmented, as an effect of the finite
size of the system in the $z$-direction. The relative strength  of the ripplon
mode is also higher for the thinner film, since the surface-to-volume ratio is
larger. One notes also that the surface mode  mantains its identity well above
the roton energy. The thicker film has a spectrum quite similar to the one of
the free surface, apart from the  2D roton at the substrate-helium interface,
which is still present but  affects the spectrum only in the region $q \simeq
2$ \AA$^{-1}$. The  spectrum is dominated by the bulk phonon-roton mode and the
lowest  surface mode,  a non negligible strength being also associated with the
excited surface modes which lie in between.

	It is worth noticing that the potential between one \hef atom and
the $H_2$ substrate is similar to that between one \hef and a Graphite
substrate coated by two solid \hef layers \cite{Cle94,Cle94b}. Hence, the
liquid \hef film on Graphite has a spatial  structure similar to  the profile
of Fig.~\ref{fig:prof_h2}.  Thus we can can safely  compare our results with
the experimental ones, obtained at ILL-Grenoble  \cite{Lau90}, as well as with
the variational theory of Ref.~\cite{Cle94b}. The spectra in
Fig.~\ref{fig:spec_h2} are indeed similar to the experimental ones (see for
example Fig.~14 of Ref.~\cite{Cle94b}). The energy of the excitations are in
rather good agreement, apart from the energy of the 2D-roton which, however,
depends crucially on the substrate-helium potential. A reasonable agreement is
also obtained for the strength distribution.

\section{Conclusions}
\label{sec:conclu}

	We have presented a new density functional theory for liquid $^4$He at
zero temperature. The theory corresponds to an improved version of the density
functional introduced in Ref.~\cite{Dup90}. It is a phenomenological theory,
where known properties of the uniform  liquid are used to fix the parameters of
the functional. The theory is suitable to study inhomogeneous states of liquid
helium in different geometries like the free surface, droplets, films, bubbles,
vortices. Equilibrium configurations and  excited states can be studied in a
unified framework. Here we have  presented results for static and dynamic
properties of the free surface and  films, as well as for the ground state  of
helium droplets. The comparison with available experimental data, as well as
with Monte Carlo  simulations, is in general satisfactory.
The present density functional
theory turns out to be quantitative even on the scale of interatomic distances.
The improvements with respect to previous density functionals are clearly
understood in term of the key ingredients of the theory, namely : i) the  bulk
equation of state, ii) the static response function and iii) the phonon-roton
dispersion in the uniform liquid. In particular we have discussed in details
the importance of reproducing the peak of the static response function at the
roton  wavelength in determining static density oscillations near the surface
and shell structures in helium droplets ("soft packing"). We have also
discussed the contribution of the collective (one-phonon) modes to sum rules
(compressibility sum rule, f-sum rule, static form factor), in order to
clarify the grounds of TDDF theory, as well as to include backflow effects. The
results for the excited states of the free surface and films compare
quantitatively  with experiments. The applications of the same theory to the
dynamics of droplets and to the phenomenon of quantum evaporation are work in
progress.

\acknowledgments
We are indebted to L. Pitaevskii for many useful discussions.
This work was partially supported by European Community grant
ERBCHRXCT920075.

%

\begin{figure}
\caption{ Static response function in liquid \hef at zero pressure. Points:
experimental data \protect \cite{Cow71}; dotted line: from functional of Refs.
\protect \cite{Str87a,Str87b}; dashed line: Orsay-Paris
functional \protect \cite{Dup90}; solid line: present functional
(Eq.~(\protect \ref{eq:chi-1}))  }
\label{fig:chi}
\end{figure}

\begin{figure}
\caption{ Pressure versus density in bulk. Points: experimental data
\protect \cite{DeB87}; dashed line: Quantum Monte Carlo results \protect
\cite{Bor94}; solid line: present density functional. }
\label{fig:eqstate}
\end{figure}

\begin{figure}
\caption{ Sound velocity in bulk. Points: experimental data \protect
\cite{Abr70}; dashed line: Quantum Monte Carlo results \protect
\cite{Bor94}; solid line: present density functional. }
\label{fig:sound}
\end{figure}

\begin{figure}
\caption{ Free surface profile of liquid \hef at zero temperature.
Dotted line: functional of Ref. \protect \cite{Str87a,Str87b};
dashed line: Orsay-Paris functional \protect \cite{Dup90};
solid line: present functional. }
\label{fig:free}
\end{figure}

\begin{figure}
\caption{ Density profile of small \hef droplets (density
normalized to the bulk value) }
\label{fig:3d}
\end{figure}

\begin{figure}
\caption{ Density profile of a droplet with $70$ \hef atoms.
Solid line: present work; triangle: DMC simulations of Ref.
\protect \cite{Chi92}; dashed line: Variational (HNC) calculations
\protect \cite{Chi94}; circles: DMC simulation of Ref. \protect
\cite{Wha94}.   }
\label{fig:clu70}
\end{figure}

\begin{figure}
\caption{Energy per particle versus $N$. Solid line: present work;
dashed line: results with functional of  Ref. \protect \cite{Dup90};
dotted line: Ref. \protect \cite{Str87b}; crosses:
Ref. \protect \cite{Mel84}; circles: Ref. \protect \cite{Chi92}. }
\label{fig:cluenergy}
\end{figure}

\begin{figure}
\caption{ Evaporation energy. Solid line: density functional; dashed
line: liquid drop formula. Circles: deviation from the liquid
drop formula (axis on the right). }
\label{fig:evap}
\end{figure}

\begin{figure}
\caption{ Evolution of $\mu$ {\it vs} $^4$He coverage on several
substrates. Solid lines: present work; dashed lines: Orsay-Paris
functional. The substrates are: H$_2$ (insert)  and, from top to bottom,
Cs, Rb, K,  Na, Li, Mg. }
\label{fig:mu_film}
\end{figure}

\begin{figure}
\caption{ Density profile of helium films on solid hydrogen from $0.02$
to $0.6$
\AA$^{-2}$. Solid lines: stable films, dotted lines: unstable or metastable
films. The growth is not continuous because of the two layering transitions
related to the formation of the first two layers.}
\label{fig:prof_h2}
\end{figure}

\begin{figure}
\caption{ Adsorption isotherm on a graphite substrate with $T = 0.639$ K. Solid
line: present work; dashed lined: Orsay-Paris  functional \protect
\cite{Cheng3}; dots: experimental results  extracted from Ref. \protect
\cite{Zim92}. One layer corresponds to $0.07812$ \AA$^{-2}$ (i.e.,
$\rho_0^{2/3}$ where $\rho_0$ is the bulk density).}
\label{fig:adsorp}
\end{figure}

\begin{figure}
\caption{ Phonon-roton dispersion in bulk at zero pressure. Points:
experimental data \protect \cite{Don81}; dotted line: Feynman
approximation (\protect \ref{eq:m1overm0}) with the experimental
static form factor $S_q$ \protect \cite{Sve80}; dashed line:
Orsay-Paris functional \protect \cite{Dup90,Pri94}; dot-dashed
line: from Eq.~(\protect \ref{eq:leastaction}) with ${\cal H}_v$
given in Eq.~(\protect \ref{eq:mv2});  solid line: same as before
but with ${\cal H}_v$ from Eq.~(\protect \ref{eq:hv}). }
\label{fig:bulk}
\end{figure}

\begin{figure}
\caption{Energy weighted moment of the dynamic structure function. Dashed
line: total f-sum rule; bars: one-phonon contribution as  measured in neutron
scattering \protect \cite{Cow71}; solid line: present work (Equation (\protect
\ref{eq:m1vj})).}
\label{fig:m1}
\end{figure}

\begin{figure}
\caption{Dispersion relation of bulk and free surface excitations.
The threshold $\Delta=8.7$ K, for roton states with different values of
parallel wave vector $k$, is shown as the horizontal line.
The present result for the surface mode (lowest solid  line
reaching $\Delta$ at $k=1.15$ \AA$^{-1}$) is compared with the
experimental data on films \protect \cite{Lau90} (squares) and
with the hydrodynamic dispersion of ripplons (short-dashed line).
The bulk phonon-roton branch (upper solid
line)  is compared with  the experimental one (circles).   The
threshold for the emission  of atoms into the vacuum is also
shown (dot-dashed curve).}
\label{fig:freesurface}
\end{figure}

\begin{figure}
\caption{Dynamic structure function (in arbitrary units) for $q_z=0$ in a
slab $50$ \AA\ thick.  The lowest line corresponds to $k=0.3$
\AA$^{-1}$; the highest one to $k=1.9$ \AA$^{-1}$. The dashed lines are the
phonon-roton and the ripplon dispersion. The $\delta$-function in the
definition (\protect \ref{eq:sqomega2}) is replaced by a gaussian
of width $0.4$ K.}
\label{fig:sqomegafree}
\end{figure}

\begin{figure}
\caption{Evolution of third sound speed $c_{3s}$ {\it versus} coverage
on a H$_2$ substrate. Solid line: Orsay-Trento results. Dashed line:
Orsay-Paris results.}
\label{fig:c3_H2}
\end{figure}

\begin{figure}
\caption{Spectrum of films on a $H_2$ substrate for coverages $0.24$
\AA$^{-2}$ (a) and $0.48$ \AA$^{-2}$ (b). Note the presence of
a low energy excitation around $2$ \AA$^{-1}$ which is associated with
a 2D roton mode trapped in the first layer.}
\label{fig:spec_h2}
\end{figure}

\begin{figure}
\caption{Dynamic structure function (in arbitrary units)
associated with the spectra in
Fig. \protect \ref{fig:spec_h2}, for coverages $0.24$ \AA$^{-2}$ (a) and
$0.48$ \AA$^{-2}$ (b) on a $H_2$ substrate. The lowest line
corresponds to $k= 0.1$ \AA$^{-1}$, the highest one to $k=2.0$  \AA$^{-1}$.
The $\delta$-function in the definition (\protect \ref{eq:sqomega2}) is
replaced by a gaussian  of width $0.4$ K. }
\label{fig:strenght_h2}
\end{figure}

\begin{table}

\caption{ Values of the potential parameters $C_3$ and $D$.}

\begin{tabular}{c|c|c|c|c|c|c|c}
 &Cs&Rb&K&Na&Li&Mg&Gr\\
\tableline
$C_3$ (K\AA$^3$)&673&754&812&1070&1360&1775&2088\\
\tableline
$D$ (K)&4.41&4.99&6.26&10.4&17.1&32.1&192.6\\
\end{tabular}
\label{tableI}
\end{table}

\begin{table}
\caption{ Values of the parameters used in $V_J(\rr)$, see Eq. (\protect
\ref{eq:vj}).}
\begin{tabular}{c|c|c|c|c|c}
$\gamma_{11}$&$\gamma_{21}$&$\gamma_{12}$&$\gamma_{22}$&$\alpha_1$&$\alpha_2$
\\
\tableline
$-19.7544$&$-0.2395$&$12.5616$ \AA$^{-2}$&$0.0312$ \AA$^{-2}$
&$1.023$ \AA$^{-2}$&$0.14912$ \AA$^{-2}$\\
\end{tabular}
\label{tableII}
\end{table}


\begin{references}
\bibitem{note:chi} We note that this definition of $\chi(q)$ differs from
the one of Ref. \protect \cite{Dup90} by a factor $\rho$.
\end{references}
\end{document}